\documentclass[12pt]{iopart}
\usepackage[dvips]{graphicx}
\usepackage{iopams}

\begin{document}
\title{Multi-fermion interaction models in curved spacetime}

\author{Masako Hayashi}
\address{Department of Physics, Hiroshima University,
Higashi-Hiroshima 739-8526, Japan}
\ead{hayashi@theo.phys.sci.hiroshima-u.ac.jp}
\author{Tomohiro Inagaki}
\address{Information Media Center, Hiroshima University,
Higashi-Hiroshima 739-8521, Japan}
\ead{inagaki@hiroshima-u.ac.jp}
\author{Hiroyuki Takata}
\address{Department of Theoretical Physics,
 Tomsk State Pedagogical University,
 Tomsk 634041, Russia}
\address{Department of Physics, Hiroshima University,
Higashi-Hiroshima 739-8526, Japan}
\ead{takataxx@tspu.edu.ru}

\begin{abstract}
A model with a scalar type eight-fermion interaction is investigated 
in curved spacetime. The ground state of the model can be obtained by
observing the effective potential. Applying the Riemann normal coordinate 
expansion, we calculate an effective potential of the model in a weakly 
curved spacetime. The result is extended to models with multi-fermion 
interactions. 
We numerically show the behavior of the effective potential and find 
the phase structure of the model.
\end{abstract}
\pacs{04.62.+v, 11.30.Qc}
\maketitle

\section{Introduction}
It is expected that a fundamental theory with a higher symmetry may 
be realized at high energy scale. The theory is inspected in critical 
phenomena at early universe. A mechanism to break the symmetry has 
played a decisive role to construct a theory of particle physics. 
One of possible mechanisms is found in hadron physics. A composite 
operator constructed by quark and anti-quark develops a non-vanishing 
expectation value and the approximate chiral symmetry is broken 
dynamically. The broken symmetry is restored at high temperature, 
density and/or strong curvature. These phenomena are understood by
non-perturbative dynamics in non-Abelian gauge theories (QCD, SM, GUT, 
$\cdots$). 

A four-fermion interaction model is introduced to particle physics by 
Nambu and Jona-Lasinio to study low energy phenomena of the strong 
interaction\cite{NJL}. The model has a similar apparent symmetry to the low 
energy limit of QCD. Non-perturbative phenomena in the model can be evaluated
trough the expansion in terms of the number of fermion flavors. It can 
be observed that the chiral symmetry is dynamically broken for a sufficiently 
strong coupling according to a vacuum condensate of the composite field
of fermions. Now some kinds of four-fermion interactions are used to 
analyze phase transition and critical behavior under extreme conditions.

There are many works to study the curvature induced phase transition in 
four-fermion interaction models. The pioneering works 
have been done by Itoyama, Buchbinder and Kirillova in two 
dimensions.\cite{Itoyama, BK}
In Refs. \cite{IMO, EOS} and \cite{Inagaki}, four-fermion interaction 
models are studied in four, three and arbitral dimensions ($2\le D<4$) 
respectively. The phase structure of the model is found in the maximally 
symmetric spacetime \cite{IMM, ELO}, Einstein space \cite{IIM} and a
curved spacetime with a nontrivial topology \cite{ELO2}.
For a review, see for example Ref. \cite{IMO2}.
The phase structure of the models depends on the coupling constants and 
the spacetime curvature. The broken chiral symmetry is restored if the 
spacetime curvature is positive and strong enough. However, the chiral 
symmetry is always broken down in a negative curvature spacetime even if 
the four-fermion coupling is small. This is one of the characteristic 
features in the four-fermion models.

Since the four-fermion interactions are the lowest dimensional operator 
to describe the fermion self interaction, we often consider only the 
four-fermion interactions at low energy.
However, it is not always valid to neglect higher dimensional operators 
in the low energy effective models.
For example, 
't~Hooft introduced a determinantal interactions in a low energy effective model of QCD to deal with the explicit breaking of the $U_A(1)$ symmetry 
\cite{tHooft}. 
R.~Alkofer and I. Zahed considered an eight-fermion interaction to explain
the pseudoscalar nonet mass spectrum \cite{Alkofer:1990uh}.
An influence of higher derivatives is considered in Ref. \cite{ELO3}.

In the present paper we consider higher dimensional scalar type interactions 
and study the curvature induced chiral symmetry breaking. 
In Sec.2 we investigate a model with four- and eight-fermion 
interactions in curved spacetime. Using the auxiliary field 
method and applying the Riemann normal coordinate expansion, we
calculate the effective potential in a weakly curved spacetime.
In Sec. 3 we introduce multi-fermion interactions and generalize
the auxiliary field method. In some specific cases we show a 
correspondence with the mean field approximation.
In Sec.4 we study the form of the effective potential near
the origin. It is shown that the chiral symmetry is always broken
in a negative curvature spacetime.
In Sec.5 we numerically evaluate the effective potential and
show the phase structure of the model with four- and eight-fermion
interactions. We discuss the contribution from the higher 
dimensional operator to the chiral symmetry breaking. Finally 
we give some concluding remarks.

\section{Eight-fermion interaction model in curved spacetime}
In this section we introduce a model with scalar type four-
and eight-fermion interactions and calculate the effective 
potential of the model in weakly curved spacetime.

Four-fermion interaction models are often used to study the 
critical phenomena of dynamical symmetry breaking. The simplest
model to induce the dynamical symmetry breaking is Gross-Neveu
model\cite{GN}. We extend the action of the Gross-Neveu model 
with an eight-fermion interaction,
\begin{eqnarray}
S&=&\int d^4 x \sqrt{-g}\left[\sum_{i=1}^{N}
 \bar{\psi}_i i\gamma^\mu(x)\nabla_\mu\psi_i \right.
\nonumber \\
 && \left. + \frac{G_1}{N} (\sum_{i=1}^{N} \bar{\psi}_i\psi_i)^2
 + \frac{G_2}{N} (\sum_{i=1}^{N} \bar{\psi}_i\psi_i)^4 \right],
\label{action}
\end{eqnarray}
where the index $i$ shows the flavors of the fermion field $\psi$,
$N$ is the number of flavors, $G_1$ and $G_2$ the coupling constants
for the four- and the eight-fermion interactions respectively, $g$ 
the determinant of the metric tensor, $\gamma^{\mu}(x)$ the Dirac
matrix in curved spacetime and $\nabla_\mu\psi$ the covariant
derivative of the fermion field $\psi$. 

The action (\ref{action}) is symmetric under the discrete chiral 
transformations,
\begin{equation}
\psi_i \rightarrow \gamma_5 \psi_i ,
\label{chiral}
\end{equation}
This chiral symmetry prevents the action from having mass term.

The action is also invariant under a global flavor transformation,
\begin{equation}
\psi_i \rightarrow \left(\exp (i\sum_a \theta_a T_a)\right)_{ij} \psi_j,
\label{flavor}
\end{equation}
where $T_a$ are generators of the flavor $SU(N)$ symmetry. The flavor
symmetry allows us to work in a scheme of the $1/N$ expansion.
Below we neglect the flavor index for simplicity.

For practical calculations it is more convenient to introduce auxiliary 
fields $\sigma_1$ and $\sigma_2$ and rewrite the action in the following
form,
\begin{equation}
S_\sigma=\int d^4x \sqrt{-g}\left[
\bar{\psi}\left(i\gamma^\mu(x)\nabla_\mu -\sigma\right)\psi 
-\frac{N \sigma_1^2}{4G_1}-\frac{N \sigma_2^2}{4G_2}
\right]\label{action2},
\end{equation}
where $\sigma$ is defined by
\begin{equation}
\sigma:=\sigma_1 \sqrt{1-\frac{N\sigma_2}{G_1}}.
\label{sigma}
\end{equation}
The equations of motion for the auxiliary fields are given by
\begin{eqnarray}
\sigma_1 &=& -\frac{2G_1}{N}\sqrt{1-\frac{N\sigma_2}{G_1}}\bar{\psi}\psi,  \label{eom1} \\
\sigma_2 &=& -\frac{2G_2}{N}(\bar{\psi}\psi)^2\;. \label{eom2}
\end{eqnarray}
Substituting these equations of motion into the action (\ref{action2}),
we obtain the original action (\ref{action}).

If the non-vanishing expectation value is assigned to $\sigma$, a mass 
term for the fermion filed $\psi$ is dynamically generated and the chiral 
symmetry is eventually broken. To study the phase structure we want to 
find the ground state of the model. In the present paper we assume that 
the spacetime curved slowly and neglect terms involving derivatives of 
the metric tensor higher than second order. We also restrict ourselves
to the static and homogeneous spacetime. In this case we can assume
that the expectation values for $\sigma_1$ and $\sigma_2$ are constant.

The energy density under the constant background $\sigma_1$ and $\sigma_2$ 
is given by the effective potential.
At the large $N$ limit the effective potential $V$ is obtained by 
evaluating vacuum bubble diagrams, 
\begin{equation}
V(\sigma_1,\sigma_2)=
\frac{\sigma_1^2}{4G_1}+\frac{\sigma_2^2}{4G_2}
+v(\sigma),
\label{v1}
\end{equation}
where we drop the over all factor $N$ and $v(\sigma)$ is given by
\begin{equation}
v(\sigma):= \frac{i}{N} \mbox{Tr}\ln\left<x|\left[
i\gamma^\mu(x)\nabla_\mu 
-\sigma
\right]|x\right>,
\end{equation}
where "Tr" denotes trace with respect to flavor, spinor indices and spacetime
coordinate. The ground state should minimize this effective potential.

Since the four- and the eight fermion interactions are nonrenormalizable, 
above $v(\sigma)$ is divergent. To obtain the finite result we regularize
the divergent integral by the cut-off method. After performing the Fourier
transformation and the Wick rotation, we introduce the four-momentum 
cut-off, $\Lambda$.
Applying the Riemann normal coordinate expansion, we expand the effective 
potential and find
\begin{eqnarray}
v(\sigma)&=&
-\frac{1}{(4\pi)^2}
\left[
\sigma^2\Lambda^2
+\Lambda^4 \ln \left(1+\frac{\sigma^2}{\Lambda^2}\right)
-\sigma^4 \ln \left(1+\frac{\Lambda^2}{\sigma^2}\right)
\right]
\nonumber \\
&&-\frac{1}{(4\pi)^2}\frac{R}{6}
\left[
-\sigma^2\ln \left(1+\frac{\Lambda^2}{\sigma^2}\right)
+\frac{\Lambda^2 \sigma^2}{\Lambda^2+\sigma^2}
\right] +O(R_{;\mu}, R^2).
\label{v11}
\end{eqnarray} 
We keep only terms independent of the curvature $R$ and terms linear 
in $R$. To determine the ground state we can freely normalize the
effective potential. Here the effective potential is normalized 
so that $V(0,0)=0$.

The necessary condition for the minimum of the effective potential is given
by the gap equations,
\begin{equation}
\frac{\partial V(\sigma_1,\sigma_2)}{\partial \sigma_1}=\frac{\sigma_1}{2G_1}
+\frac{dv(\sigma)}{d\sigma}\frac{\partial \sigma}{\partial \sigma_1}=0,
\label{gapA}
\end{equation}
and
\begin{equation}
\frac{\partial V(\sigma_1,\sigma_2)}{\partial \sigma_2}=\frac{\sigma_2}{2G_2}
+\frac{dv(\sigma)}{d\sigma}\frac{\partial \sigma}{\partial \sigma_2}=0,
\label{gapB}
\end{equation}
with
\begin{eqnarray}
\frac{\partial \sigma}{\partial \sigma_1}&=&
\sqrt{1-\frac{N\sigma_2}{G_1}}\;,\;\;
\frac{\partial \sigma}{\partial \sigma_2}=-\frac{N\sigma_1}{
2G_1\sqrt{1-\frac{\displaystyle N \sigma_2}{\displaystyle G_1}}},
\nonumber \\
\frac{dv(\sigma)}{d\sigma}&=&-\frac{\sigma}{4\pi^2}
\left[
\Lambda^2-\sigma^2 \ln \left(1+\frac{\Lambda^2}{\sigma^2}\right)
\right]
\nonumber \\
&&-\frac{\sigma}{48\pi^2}R
\left[
-\ln \left(1+\frac{\Lambda^2}{\sigma^2}\right)
+ \frac{\Lambda^2}{\Lambda^2+\sigma^2}
+ \frac{\Lambda^4}{(\Lambda^2+\sigma^2)^2}
\right].
\end{eqnarray}

Under the ground state the expectation values for $\sigma_1$ and 
$\sigma_2$ satisfy these gap equations.
From the gap equations (\ref{gapA}) and (\ref{gapB}) we find the 
following conditions for non-vanishing expectation values of 
$\sigma_1$ and $\sigma_2$.
\begin{equation}
\langle\sigma_1\rangle^2
=\frac{2G_1}{G_2}
\langle\sigma_2\rangle
\left(\langle\sigma_2\rangle-\frac{G_1}{N}\right), 
\label{hyperbolic}
\end{equation}
or
\begin{eqnarray}
\frac{1}{2G_1}
-\frac{1}{4\pi^2}
\left[
\Lambda^2-\langle\sigma_1\rangle^2 \ln \left(1+\frac{\Lambda^2}{\langle\sigma_1\rangle^2}\right)
\right]
\nonumber \\
-\frac{R}{48\pi^2}
\left[
-\ln \left(1+\frac{\Lambda^2}{\langle\sigma_1\rangle^2}\right)
+ \frac{\Lambda^2}{\Lambda^2+\langle\sigma_1\rangle^2}
+ \frac{\Lambda^4}{(\Lambda^2+\langle\sigma_1\rangle^2)^2}
\right]=0,
\nonumber \\
\langle\sigma_2\rangle = 0.
\label{fourfermitype}
\end{eqnarray}
One of the conditions (\ref{hyperbolic}) and (\ref{fourfermitype}) should be 
satisfied at the minimum of the effective 
potential. In the former case the expectation values for $\sigma_1$ and 
$\sigma_2$ are on a hyperbolic curve for a positive $G_1 G_2$ or on an 
ellipse for a negative $G_1G_2$. The latter condition is the same as that
in the four-fermion interaction model and the eight-fermion interaction has 
nothing to do with the ground state.

At the ground state the expectation values obey the classical equations 
of motion.
Substituting Eqs.(\ref{eom1}) and (\ref{eom2}) into the conditions
(\ref{hyperbolic}) or (\ref{fourfermitype}), we find one of the following 
relationships,
\begin{equation}
\left<(\bar{\psi}\psi)^2\right>=\left<(\bar{\psi}\psi)\right>^2,
\mbox{ or }
\left<(\bar{\psi}\psi)^2\right>=0.
\end{equation}
These relationships mean that our result at the leading order of the 
$1/N$ expansion coincides with the one in the mean field approximation
except for the case, $\left<(\bar{\psi}\psi)^2\right>=0$. 

We numerically evaluate the effective potential (\ref{v1}) and show the phase 
structure of the model in Sec.4.

\section{Multi-fermion interactions}
The results in the previous section can be extended more general cases.
The extension of the model is not unique. Here we assume that vector, tensor 
type interactions and interactions with derivative operator do not
develop expectation values at the ground state and contribution to
the phase structure is negligible. We consider models with only scalar-type 
multi-fermion interactions and calculate the gap equation.

We start from the action defined by 
\begin{equation}
S=\int d^4x \sqrt{-g}\left[\bar{\psi}i\gamma^\mu(x)\nabla_\mu \psi 
+ \sum^{n}_{k=1} G_k (\bar{\psi}\psi)^{2k} \right].
\label{action:mul}
\end{equation}
This action is invariant under the discrete chiral transformation
(\ref{chiral}) and the global flavor transformation (\ref{flavor}).

According to the functional integral formalism, the generating
functional is given by
\begin{equation}
Z=\int {\cal D}\tilde{\psi}{\cal D}\tilde{\bar{\psi}}e^{iS},
\label{gen1}
\end{equation}
where we set the path-integral measure, 
$\tilde{\psi} \equiv \sqrt[4]{-g}\psi$ and 
$\tilde{\bar{\psi}} \equiv \sqrt[4]{-g}\bar{\psi}$, 
to keep the general covariance. 

We consider a Gaussian integral
\begin{eqnarray}
C&:=&
\int \prod_{k=1}^n {\cal D}\sigma_k
\nonumber \\
&& \times
\exp\left\{
i\int d^4x \sqrt{-g}\sum_{l=1}^n
\left[
-\frac{N}{4G_l}
\left(
\sigma_l+\frac{1}{N}\sum^l_{m=1}a_{lm}(\bar{\psi} \psi)^m
\right)^2
\right]
\right\},
\end{eqnarray} 
and inserts it in the right-hand side of Eq.(\ref{gen1}). 
Thus the generating functional (\ref{gen1}) is 
rewritten as 
\begin{equation}
Z=\frac{1}{C}\int{\cal D}\tilde{\psi}{\cal D}\tilde{\bar{\psi}}
\prod_{k=1}^n {\cal D}\sigma_k e^{iS_y},
\end{equation}
where the action $S_y$ is given by
\begin{eqnarray}
S_y&=&\int d^4x\sqrt{-g}
\left[
\bar{\psi}i\gamma^\mu \nabla_\mu \psi
+\sum_{k=1}^n\frac{G_k}{N}(\bar{\psi}\psi)^{2k}
\right.
\nonumber \\
&&\left.
-\sum_{k=1}^n\frac{N}{4G_k}\left(\sigma_k
+\frac{1}{N}\sum_{l=1}^k a_{kl}(\bar{\psi}\psi)^l\right)^2
\right].
\label{sy1}
\end{eqnarray}
Since the number of arbitrary parameters, $a_{kl}$, is $n(n+1)/2$,
we can choose
$a_{kl}=a_{kl}(\sigma_{k+1}, \sigma_{k+2}, \cdots , \sigma_n)$ 
to satisfy
\begin{eqnarray}
\sum_{k=1}^n \frac{N}{4G_k} 
\left(
\sigma_k +\frac{1}{N}\sum_{l=1}^k a_{kl}(\bar{\psi}\psi)^l
\right)^2
\nonumber\\
=f(\sigma)\bar{\psi}\psi
+ \sum_{k=1}^n 
\left(
\frac{G_k}{N}(\bar{\psi}\psi)^{2k} 
+\frac{N}{4G_k}\sigma_k^2
\right),
\label{cond1}
\end{eqnarray}
where $f(\sigma)$ is a function of $\sigma_k$ $(k=1,2,\cdots,n)$.
Therefore the multi-fermion interaction terms in Eq.(\ref{sy1}) are
canceled out and the action $S_y$ reduces to
\begin{equation}
S_y=\int d^4x\sqrt{-g}
\left[
\bar{\psi}(i\gamma^\mu \nabla_\mu -f(\sigma))\psi
-\sum_{k=1}^n\frac{N}{4G_k}{\sigma_k}^2
\right].
\label{sy2}
\end{equation}
Applying the similar method mentioned in the previous section,
we obtain the effective potential
\begin{equation}
V(\sigma_1,\cdots,\sigma_n)=\sum_{k=1}^n
\frac{{\sigma_k}^2}{4G_k}+v(f(\sigma)),
\label{v2}
\end{equation}
where $v$ is defined in Eq.(\ref{v11}).

We calculate the explicit expression of $f(\sigma)$ for some cases.
For $n=2$ it is easy to reproduce the result in Eq. (\ref{sigma}).
As another simple case, we consider the action,
\begin{equation}
S=\int d^4x \sqrt{-g}\left[\bar{\psi}i\gamma^\mu(x)\nabla_\mu \psi
+ \sum^{n}_{k=1} \frac{g_k}{N} (\bar{\psi}\psi)^{2^k} \right].
\label{ac:mul}
\end{equation}
In this case the function $f(\sigma)$ is given by the following 
induction series,
\begin{equation}
f(\sigma):=\sigma_{1}^\prime, 
\sigma_{k}^\prime:= \sigma_{k}\sqrt{1-\frac{N\sigma_{k+1}^\prime}{g_{k}}}\;\; (k=1,\cdots n) ,
\sigma_{n+1}^\prime:=0.
\end{equation}
It is straightforward to show that above $f(\sigma)$ satisfies,
\begin{eqnarray}
\sum_{k=1}^n \frac{N}{4g_k} 
\left(
\sigma_k +\frac{1}{N} a_{k}(\bar{\psi}\psi)^{2^{k-1}}
\right)^2
\nonumber\\
=f(\sigma)\bar{\psi}\psi
+ \sum_{k=1}^n 
\left(
\frac{g_k}{N}(\bar{\psi}\psi)^{2^k} 
+\frac{N}{4g_k}\sigma_k^2
\right),
\label{cond2}
\end{eqnarray}
with
\begin{equation}
a_k:=2g_k\sqrt{1-\frac{N\sigma_{k+1}^\prime}{g_k}} ,
a_n:=-2g_n.
\end{equation}
for $k=1,\cdots n$.
It is a sufficient condition for Eq.(\ref{cond1}). 
Hence, the action (\ref{ac:mul}) is simplifies to
\begin{equation}
S_y=\int d^4x\sqrt{-g}
\left[
\bar{\psi}(i\gamma^\mu \nabla_\mu -f(\sigma))\psi
-\sum_{k=1}^n\frac{N}{4g_k}{\sigma_k}^2
\right].
\label{sy3}
\end{equation}

Since the effective potential has the same form with Eq.(\ref{v2}),
the gap equations for the action (\ref{sy3}) are given by
\begin{equation}
\frac{\partial V(\sigma_1,\cdots,\sigma_n)}{\partial \sigma_k}=\frac{\sigma_k}{2g_k}
+\frac{dv(f(\sigma))}{df(\sigma)}\frac{\partial f(\sigma)}{\partial \sigma_k}=0\;,\;\;(k=1,2,\cdots,n)
\label{gap:mul}
\end{equation}
Differentiating $f(\sigma)$ with respect to $\sigma_k$, we get
\begin{equation}
\frac{\partial f(\sigma)}{\partial \sigma_{k}}
=\frac{\partial \sigma}{\partial \sigma_{1}^\prime}
\frac{\partial \sigma_1^\prime}{\partial \sigma_{2}^\prime}
\cdots
\frac{\partial \sigma_{k-1}^\prime}{\partial \sigma_{k}^\prime}
\frac{\partial \sigma_{k}^\prime}{\partial \sigma_{k}}
= \prod_{i=1}^{k-1}
\sigma_i\frac{-N/g_i}{2\sqrt{1-\frac{N\sigma_{i+1}^\prime}{g_i}}}\sqrt{1-\frac{N\sigma_{k+1}^\prime}{g_k}},
\label{gap:mul2}
\end{equation}
for $k=1,2,\cdots,n$. We calculate the ratio between $(k-1)$-th and 
$k$-th equations of (\ref{gap:mul}),
\begin{equation}
\frac{g_{k-1}}{g_{k}} \frac{\sigma_{k}}{\sigma_{k-1}}
=
\left(\frac{\partial f(\sigma)}{\partial \sigma_{k}}\right)
\bigg/
\left(\frac{\partial f(\sigma)}{\partial \sigma_{k-1}}\right).
\label{eq:rel_k_k-1}
\end{equation}
Substituting the Eq.(\ref{gap:mul2}) into Eq.(\ref{eq:rel_k_k-1}), 
we obtain the following configuration for $\sigma_k$ at the ground state,
\begin{equation}
\frac{\sigma_k}{2g_k}\frac{1}{\sqrt{1-\frac{N \sigma_{k+1}^\prime}{g_k}}}
=-\frac{N \sigma_{k-1}^2}{4 g_{k-1}^2} 
\frac{1}{1-\frac{N \sigma_k^\prime}{g_{k-1}}},
\mbox{ or } \sigma_k=0.
\label{gap2}
\end{equation}

For the action (\ref{sy3}) the equations of motion are given by
\begin{equation}
\sigma_{k}=-\frac{2g_{k}}{N}
\sqrt{1-\frac{N \sigma_{k+1}^\prime}{g_{k}}}(\bar{\psi}\psi)^{2^{k-1}}.
\label{eom3}
\end{equation}
At the ground state it is valid to insert Eq.(\ref{eom3}) into Eq.(\ref{gap2}).
Then we eliminate $\sigma_{k-1}$, $\sigma_{k}$, $\sigma_{k}^\prime$ and 
$\sigma_{k+1}^\prime$ from both the sides in 
Eq.(\ref{gap2}) and find following relationships between expectation values,
\begin{equation}
\langle (\bar{\psi}\psi)^{2^k}\rangle =
\langle \bar{\psi}\psi \rangle ^{2^k},
\mbox{ and } \langle (\bar{\psi}\psi)^{2^{k'}}\rangle =0.
\end{equation}
for $k=1, 2, \cdots, k'$ and $k'\in \{2, \cdots, n+1\}$.
The number $k'$ depends on details of the effective potential and is
fixed by observing the minimum of that.
Therefore the solution of the gap equations (\ref{gap:mul}) are equivalent to 
the ones obtained by the mean fields approximation for $k<k'$.

\section{Chiral symmetry breaking in a negative curvature spacetime}

In a scalar type four-fermion interaction model the chiral symmetry 
is always broken in a negative curvature spacetime. Here we study the 
behavior of the effective potential near the origin and show the chiral 
symmetry breaking for $R<0$ in a multi-fermion interaction model.

First we consider the eight-fermion interaction model (\ref{action}).
To obtain a precise form of the effective potential we evaluate the
curvature of the effective potential at the limit
$\sigma_1\rightarrow 0$ and $\sigma_2\rightarrow 0$.
Differentiating the effective potential (\ref{v1}) twice
and using the condition (\ref{cond1}), we get
\begin{equation}
\frac{\partial^2 V}{\partial \sigma_1^2}=\frac{1}{2G_1}
+\left(1-\frac{N\sigma_2}{G_1}\right)\frac{d^2 v}{d\sigma^2},
\label{d2vs1}
\end{equation}
and
\begin{equation}
\frac{\partial^2 V}{\partial \sigma_2^2}=\frac{1}{2G_2}
-\frac{N\sigma_2}{2 G_2}
\left(\frac{1}{\sigma}\frac{d v}{d\sigma}+\frac{d^2 v}{d\sigma^2}\right).
\label{d2vs2}
\end{equation}
For a large $\sigma_i$ the effective potential (\ref{v1}) is almost proportional
to $\sigma_i^2$. Thus the potential has a stable minimum. 
If $\partial^2 V/\partial \sigma_1^2$ or 
$\partial^2 V/\partial \sigma_1^2$ is negative at the origin, 
the effective potential is unstable and the chiral symmetry has to
be broken.
Taking the limit $\sigma_1\rightarrow 0$ and $\sigma_2\rightarrow 0$, 
we obtain,
\begin{equation}
\frac{\partial^2 V}{\partial \sigma_1^2}\rightarrow
\lim_{\sigma\rightarrow 0}\left[
\frac{1}{2G_1}-\frac{\Lambda^2}{4\pi^2}+\frac{R}{4\pi^2}
\left(\mbox{ln}\left(1+\frac{\Lambda^2}{\sigma^2}\right)-2\right)
\right],
\label{lim:d2vs1}
\end{equation}
and
\begin{equation}
\frac{\partial^2 V}{\partial \sigma_2^2}\rightarrow\frac{1}{2G_2}.
\label{lim:d2vs2}
\end{equation}
The right-hand side in Eq.(\ref{lim:d2vs1}) is always negative for $R<0$.
It implies that the auxiliary fields, $\sigma_1$ and $\sigma_2$, develops 
non-vanishing values at the minimum of the effective potential.
Thus only the broken phase realizes in a negative curvature spacetime.

Next we consider the multi-fermion interaction model (\ref{action:mul}).
Differentiating both the sides of Eq.(\ref{cond1}) in terms of 
$\sigma_1$, we obtain
\begin{equation}
  \frac{\partial f(\sigma)}{\partial \sigma_1}
  =\frac{1}{2G_1}a_{11}(\sigma_2,\sigma_3,\cdots,\sigma_n).
\label{cond:pr1}
\end{equation}
It should be noted that that the function $a_{11}$ does not depend 
on $\sigma_1$ by definition.
Using the condition (\ref{cond:pr1}), we differentiate the effective 
potential (\ref{v2}) twice in terms of $\sigma_1$ and get
\begin{equation}
  \frac{\partial^2 V}{\partial {\sigma_1}^2}=
  \frac{1}{2G_1}+\frac{{a_{11}}^2}{4{G_1}^2}\frac{d^2 v}{d\sigma^2}.
\label{cond:pr2}
\end{equation}
To find the behavior of the effective potential near the origin, we take 
the limit $\sigma_2, \sigma_3, \cdots, \sigma_n \rightarrow 0$. At the limit 
Eq.(\ref{cond1}) we obtain,
\begin{equation}
  a_{11}(\sigma_2, \sigma_3, \cdots, \sigma_n \rightarrow 0) = 2G_1.
\label{a11}
\end{equation}
Taking the same limit in Eq.(\ref{cond:pr2}) and substituting 
Eq.(\ref{a11}), we obtain Eq.(\ref{lim:d2vs1}) again.
According to the same argument as the eight-fermion interaction model
at least one of the auxiliary fields, $\sigma_1$, develops a non-vanishing 
value at the minimum of the effective potential.
Therefore we conclude that only the broken phase realizes for the scalar
type multi-fermion interaction models in a negative curvature spacetime.

\section{Numerical analysis of the phase structures}
Here we study the contribution from higher dimensional operators to 
chiral symmetry breaking in curved spacetime. We consider the model
with four- and eight-fermion interactions and numerically evaluate 
the effective potential (\ref{v1}) with (\ref{v11}). We normalize
the mass scale by the cut-off scale $\Lambda$.
The auxiliary field $\sigma$ plays a role as an order parameter for 
the chiral symmetry breaking.

We start to calculate the effective potential with fixed coupling
constants $G_1$ and $G_2$. For $G_2=0$ the model contains only the 
four-fermion interaction. In this case the chiral symmetry is broken
down for $G_1\Lambda^2 > 2\pi^2\equiv G^{(R=0)}_{1cr}$ in Mikowski 
spacetime, i.e. R=0. In a negative curvature spacetime we observe
only the broken phase.\cite{IMO}

For a model with a finite $G_2$ we numerically evaluate the effective 
potential under the constraints (\ref{hyperbolic}) and (\ref{fourfermitype}).
We observe that the minimum of the effective potential satisfies the 
constraint (\ref{hyperbolic}) in all cases we consider here. 
Therefore we only show the results under  the constraint (\ref{hyperbolic}) 
below.

\begin{figure}[htbp]
 \begin{minipage}{0.495\hsize}
  \begin{center}
   \includegraphics[width=76mm]{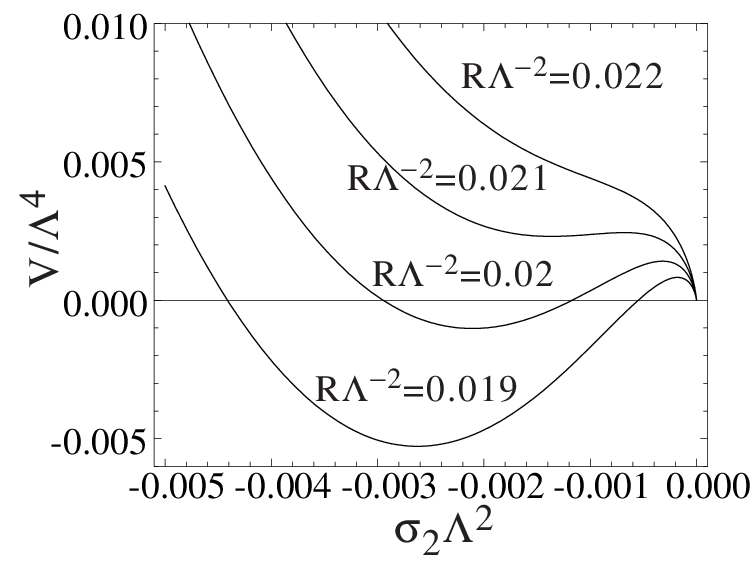}
  \end{center}
   \caption{{\footnotesize Behavior of the effective potential (\ref{v1}) for $G_1\Lambda^2=20$ and 
$G_2\Lambda^8=1000$.}}
  \label{fig:1}
  \end{minipage}
 \begin{minipage}{0.495\hsize}
  \begin{center}
   \includegraphics[width=76mm]{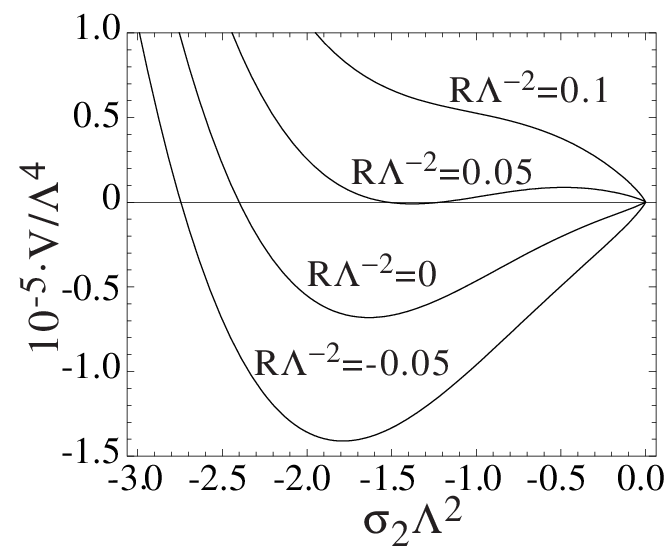}
  \end{center}
  \caption{{\footnotesize Behavior of the effective potential (\ref{v1}) for
  $G_1\Lambda^2=20$ and 
$G_2\Lambda^8=8500$.}}
  \label{fig:2}
 \end{minipage}
\end{figure}

First we consider the model with a four-fermion coupling,
$G_1 ( > G^{(R=0)}_{1cr})$, larger than the critical one for $R=0$
and $G_2=0$. 
In Figs.\ref{fig:1} and \ref{fig:2} we draw the behavior of the 
effective potential as a function of $\sigma_2$ 
for $G_1\Lambda^2=20$, $G_2\Lambda^8=1000$ and 
$G_1\Lambda^2=20$, $G_2\Lambda^8=8500$ 
respectively.
In Fig.\ref{fig:1} the chiral symmetry is broken for a negative 
curvature and is restored through the first order phase transition
at $R=R_{cr}$, as the curvature increases. Then the symmetric 
phase is realized for $R > R_{cr}$. 
In Fig.\ref{fig:2} we plot the effective potential for a larger
eight-fermion coupling, $G_2$. As is seen in the figure, the broken 
chiral symmetry is restored through the first order phase transition.
The critical curvature for $G_2\Lambda^8=8500$ is larger than 
that for $G_2\Lambda^8=1000$. Thus the chiral symmetry breaking 
is enhanced by the eight-fermion interaction.

\begin{figure}[htbp]
\begin{minipage}{0.495\hsize}
  \begin{center}
   \includegraphics[width=76mm]{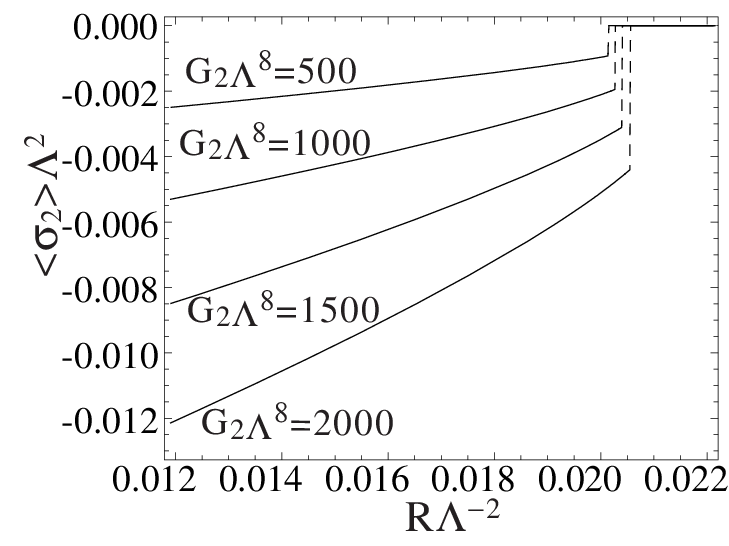}
  \end{center}
  \caption{{\footnotesize Behavior of the dynamically generated fermion mass
  for $G_1\Lambda^2=20$ and $G_2\Lambda^8=500\sim 2000$.}}
  \label{fig:3}
 \end{minipage}
 \begin{minipage}{0.495\hsize}
  \begin{center}
   \includegraphics[width=76mm]{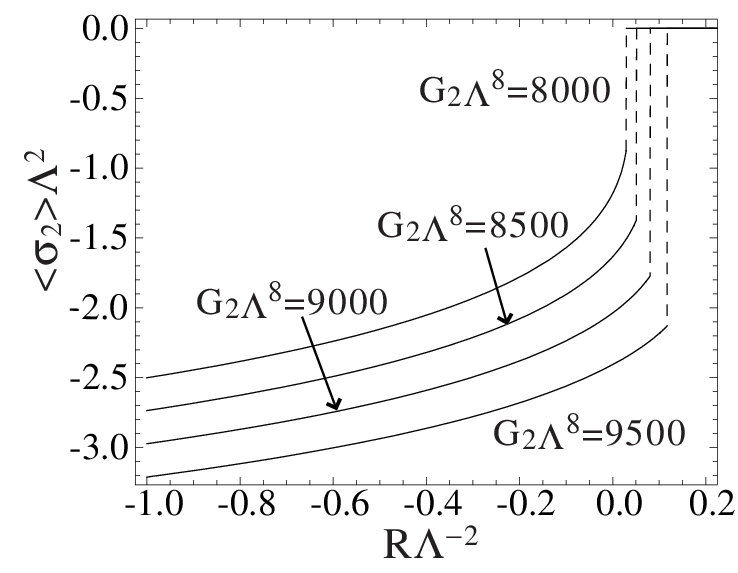}
  \end{center}
  \caption{{\footnotesize Behavior of the dynamically generated fermion mass
  for $G_1\Lambda^2=20$ and $G_2\Lambda^8=8000\sim 9500$.}}
  \label{fig:4}
 \end{minipage}
 \end{figure}

Observing the minimum of the effective potential, we obtain the 
expectation value $\langle\sigma_2\rangle$ at the ground state.
The expectation value satisfies the gap equations. A non-vanishing
$\langle\sigma_1\rangle$ is obtained for a non-vanishing and a negative 
$\langle\sigma_2\rangle$ from the constraint (\ref{hyperbolic}).
It implies a non-vanishing $\langle\sigma\rangle$ which corresponds
to the dynamically generated fermion mass. Therefore the chiral symmetry
is broken down for a non-vanishing and a negative $\langle\sigma_2\rangle$.
On the other hand the expectation value $\langle\sigma\rangle$ is vanish
for $\langle\sigma_2\rangle=0$. In this case the dynamical fermion mass 
is not generated and the chiral symmetry is guaranteed. Therefore we can
inspect the chiral symmetry breaking by the expectation value 
$\langle\sigma_2\rangle$.

In Figs.\ref{fig:3} and \ref{fig:4} we draw the expectation value 
$\langle\sigma_2\rangle$
at $G_1\Lambda^2=20$ as a function of the spacetime curvature, $R$.
It is clearly seen that the behavior of each lines are consistent with 
Figs.\ref{fig:1} and \ref{fig:2}.
The expectation value develops a non-vanishing value for a smaller 
$R$ and disappears, as the curvature increases.
We observe a mass gap in Figs.\ref{fig:3} and \ref{fig:4} which reflects 
the nature of the first order phase transition.

\begin{figure}[htbp]
\begin{minipage}{0.495\hsize}
  \begin{center}
   \includegraphics[width=76mm]{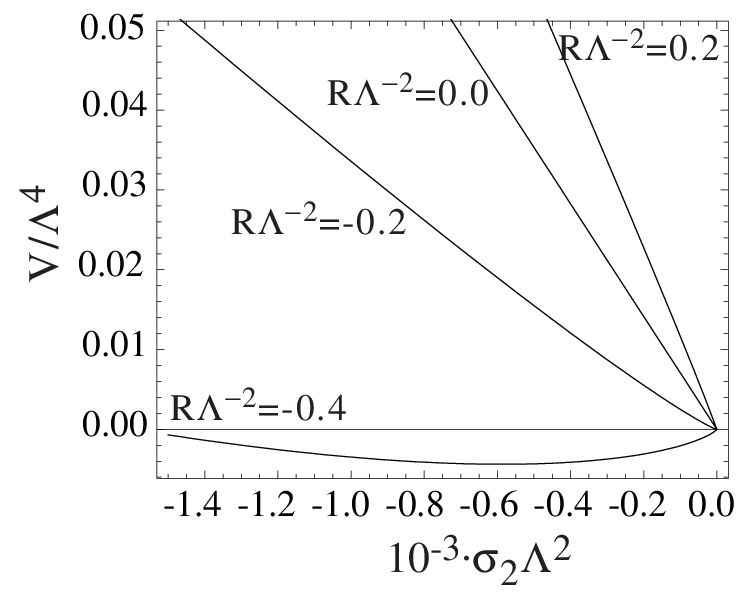}
   \end{center}
  \caption{{\footnotesize Behavior of the effective potential (\ref{v1}) for
 $G_1\Lambda^2=15$ and
  $G_2\Lambda^8=8500$.}}
  \label{fig:5}
  \end{minipage}
 \begin{minipage}{0.495\hsize}
  \begin{center}
   \includegraphics[width=76mm]{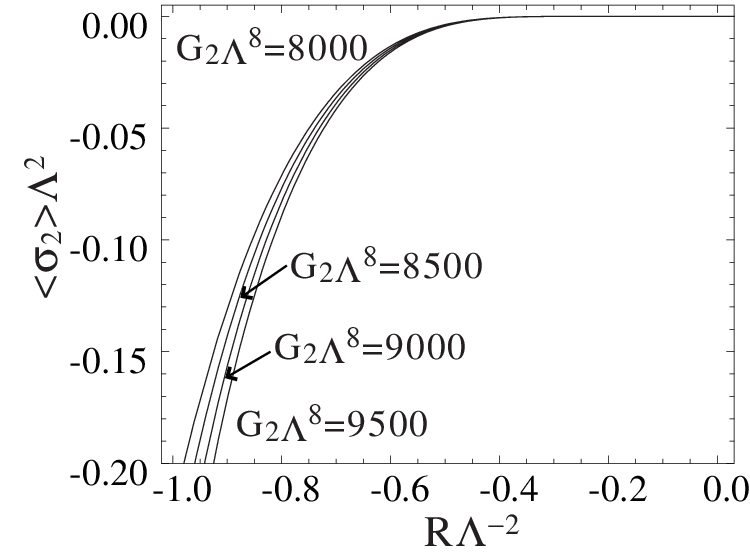}
  \end{center}
  \caption{{\footnotesize Behavior of the dynamically generated fermion mass
  for $G_1\Lambda^2=15$.}}
  \label{fig:6}
 \end{minipage}
\end{figure}

Next we evaluate the model with a four-fermion coupling,
$G_1 ( < G^{(R=0)}_{1cr})$, smaller than the critical one for $R=0$
and $G_2=0$. 
An interesting behavior is found for a smaller $G_1\Lambda^2$.
At a glance we may observe a symmetric phase even for a negative 
curvature spacetime at $G_1\Lambda^2=15$ in Fig.\ref{fig:5}.

To find the precise behavior of the effective potential we solve the gap
equation.
In Fig.\ref{fig:6} we plot the solution of the gap equation as a function
of the curvature.
We observe that the expectation value, $\langle\sigma_2\rangle$, smoothly 
disappears as the curvature increases. 
The transition from the broken phase to the symmetric phase is crossover.
Therefore it is difficult to find the critical curvature of these figures. 
As is shown in Sec.4, the expectation value $\langle\sigma_2\rangle$ 
develops a non-vanishing value and the chiral symmetry is always broken
for $R<0$.

\begin{figure}[htbp]
 \begin{minipage}{0.495\hsize}
  \begin{center}
   \includegraphics[width=76mm]{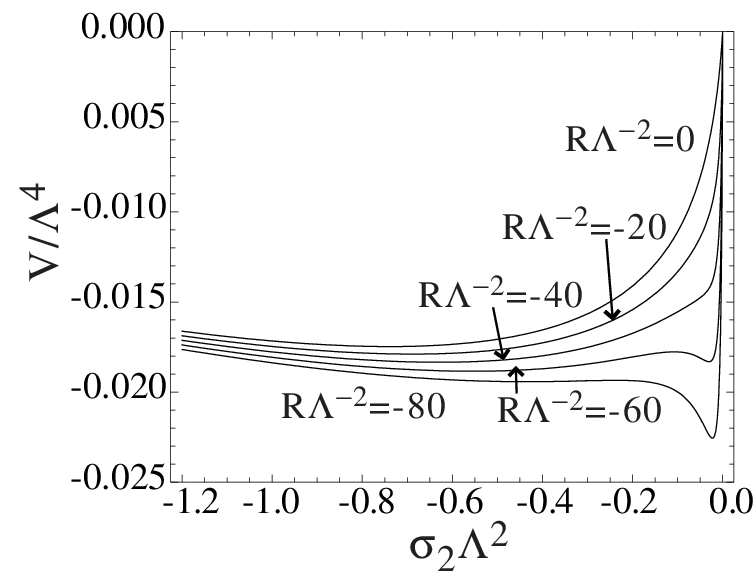}
   \end{center}
  \caption{{\footnotesize Behavior of the effective potential (\ref{v1}) for
  $G_1\Lambda^2=1000$ and
 $G_2\Lambda^8=20000$.}}
  \label{fig:7}
  \end{minipage}
 \begin{minipage}{0.495\hsize}
  \begin{center}
   \includegraphics[width=76mm]{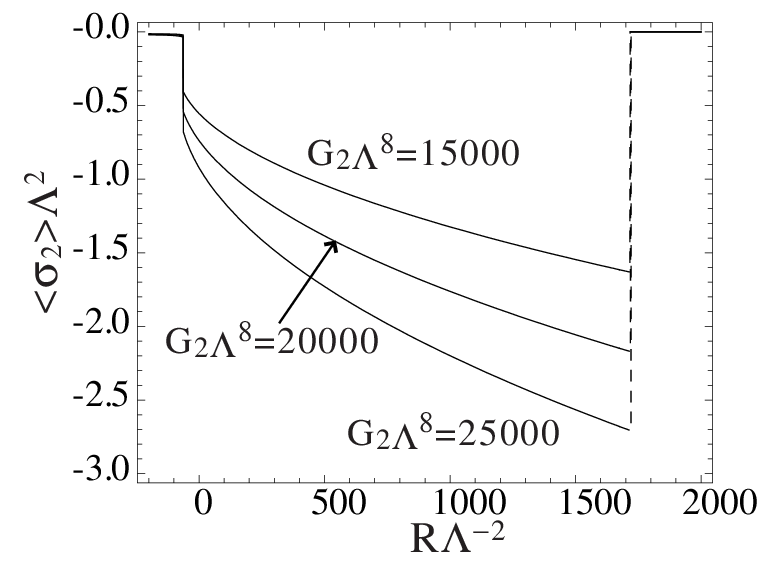}
  \end{center}
  \caption{{\footnotesize Behavior of the dynamically generated fermion mass
  for $G_1\Lambda^2=1000$.}}
  \label{fig:8}
 \end{minipage}
\end{figure}

We also consider much larger $G_1\Lambda^2$. In Fig.\ref{fig:7}
the effective potential is drawn for $G_1\Lambda^2=1000$ and
 $G_2\Lambda^8=20000$, although the coupling constant $G_1$
seems to be too large as a realistic model. It should be noted that we 
do not use a perturbative expansion in terms of the coupling constant.
We observe four extrema at $R/\Lambda^2=-60$. 
As is shown in Fig.\ref{fig:8}, we find the two kinds of the gaps for 
$\langle\sigma_2\rangle$ at a negative $R$ and a positive $R$. 
As increasing the curvature, $R$, the chiral symmetry breaking
is much enhanced at the gap with a negative $R$ and the broken 
chiral symmetry is restored through the first order phase transition 
at the gap with a positive $R$. 
The gaps appear at a large curvature $|R/(\Lambda^2)|>1$.
Since we work in the Riemann normal coordinate expansion and keep only
terms up to linear in $R$, we may spoil the validity of the expansion.
These types of the phase transition  in Figs. \ref{fig:7} and \ref{fig:8}  may
be modified by a higher order correction about $R$.  

\begin{figure}[htbp]
 \begin{minipage}{0.495\hsize}
  \begin{center}
   \includegraphics[width=76mm]{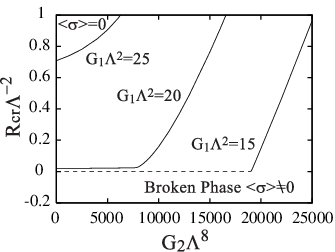}
  \end{center}
  \caption{{\footnotesize Phase diagram of the eight-fermion interaction model (\ref{action})
  for $G_1\Lambda^2=15, 20$ and $25$.}}
  \label{fig:9}
 \end{minipage}
\end{figure}

By observing the behavior of $\langle\sigma_2\rangle$, we obtain
the phase boundary which separates the chiral symmetric and the broken 
phases. The boundary curves are shown in Fig.\ref{fig:9}. Below the lines 
the chiral symmetry is broken.
The first order phase transition is shown by solid lines. The crossover boundary
is drown by dashed lines. 
We find that the chiral symmetry is broken down for an eight-fermion coupling
large enough. In the case of the crossover the critical value of the curvature, 
$R_{\mbox{cr}}$, is almost independent on the eight-fermion coupling, $G_2$.  
If the phase transition is of the first order, the chiral symmetry breaking is 
enhanced for a larger eight-fermion coupling  $G_2$.

In this section we confine our selves in the eight-fermion interaction model
and numerically evaluate the effective potential and solve the gap equation.
It is possible to apply the procedure used here to the multi-fermion interaction
models. Since the multi-fermion interaction models have  more parameters,
we expect that a more complex phase structure is realizes. Before analyze
the phase structure more general models with many parameters, it is better 
to fix the physical phenomena to apply the model and reduce the parameters. 

\section{Conclusions}
We have considered the multi-fermion interaction models as effective models to 
describe dynamical symmetry breaking at high energy scale and investigated the 
phase structure in weakly curved spacetime. 
Applying the Riemann normal coordinate expansion, we have obtained the explicit 
expression for the effective potential in the leading order of the $1/N$ expansion.
One of the solutions of the gap equation coincides with that in the mean field 
approximation. 

To inspect the existence of a non-trivial ground state we have analytically evaluated
the second derivative of the effective potential at the trivial solution of the gap equation, 
$\sigma_1=\sigma_2=0$.  The result is always negative for $R<0$. It implies that
the state with $\sigma_1=\sigma_2=0$ is unstable. Thus only the broken phase 
can be realized in a negative curvature spacetime. We conclude that the chiral 
symmetry is always broken in a negative curvature spacetime at the large $N$ limit . 
It is one of the characteristic features of the multi-fermion interaction model.

We have numerically evaluated details of the phase structure for the eight-fermion
interaction model. The broken chiral symmetry was restored at a certain critical
curvature, $R_{cr}(\geq 0)$. 
The contribution of the eight-fermion coupling enhances the chiral symmetry 
breaking.
For a small four-fermion coupling, $G_1 < G^{(R=0)}_{1cr}$, we observe the crossover
behavior, i.e. the non-vanishing expectation value, $\langle \sigma\rangle$, smoothly 
disappears, as the curvature $R$ increases.
For $G_1 > G^{(R=0)}_{1cr}$ the mass gap appears for the solution of the gap equation
and only the first order phase transition realized.
We found the phase boundary dividing  the symmetric phase and the broken phase.
There is no symmetric phase for a negative $R$, as is shown analytically.

We also apply our result to a strongly curve spacetime, though we may spoil the 
validity of the weak curvature expansion. (For the four-fermion interaction model
the validity of the weak curvature expansion in the strongly curved spacetime is 
discussed in Ref.\cite{Inagaki} .) 
A more complex phase structure is found 
in this case. We have observed two local minima for the effective potential and two 
kinds of mass gap for a negative and a large positive curvature.

We are interested in applying our results to critical phenomena in the early universe. 
A supersymmetric extension of the four-fermion interaction model is 
discussed in Ref.{BIO}. A contribution from supersymmetric partners may play an 
important role at high energy scale.
We will continue our work further and hope  to report on these problems.

\section*{Acknowledgements}
The authors would like to thank Y. Mizutani for fruitful discussions. 
T. I. is supported by the Ministry of Education, Science, Sports and 
Culture, Grant-in-Aid for Scientific Research (C), No. 18540276,
2008.


\end{document}